\documentclass{moriond}

\usepackage{ifthen}

\newcommand{\bq}{\begin{eqnarray}}
\newcommand{\eq}{\end{eqnarray}}
\newcommand{\eps}{\varepsilon}

\newcommand{\Photo}{}

\newboolean{arxiveversion}
\setboolean{arxiveversion}{true}

\begin{document}
\ifthenelse{\boolean{arxiveversion}}
{
\hfill MITP/15-032\\
\vspace*{35mm}
}
{
\vspace*{4cm}
}
\title{Precision on the top mass}

\author{ Stefan Weinzierl }

\address{PRISMA Cluster of Excellence, Institut f{\"u}r Physik, Johannes Gutenberg-Universit{\"a}t Mainz, \\
D - 55099 Mainz, Germany \vspace*{4mm} }

\maketitle
\abstracts{
In this talk I will focus on theoretical issues related to high precision determinations of the top mass.
Several mass definitions are reviewed and their respective advantages and disadvantages are discussed.
Precision determinations of the top mass will require a short-distance mass definition.
I will summarise current work in this direction.
}

% -------------------------------------------------------------------------------------------
\section{Introduction}

The top quark mass -- or equivalently the top Yukawa coupling -- is one of the fundamental parameters 
of the Standard Model.
Precise values for these fundamental parameters encode our current knowledge of the Standard Model
and are required for various reasons.
Focusing on the top quark mass, the motivations for a precision determination are as follows:
First of all, the value of the top mass affects the theory predictions for top quark cross sections.
It is therefore relevant in comparing measured top quark cross sections from the Tevatron and the ongoing LHC experiments
with theoretical predictions of the Standard Model.
Secondly, the value of the top mass affects searches for new particles in beyond the Standard Model (BSM) scenarios.
Examples are searches for processes with top background or BSM decays into top quarks.
For these first two reasons it is desirable to determine the top quark mass at least to a precision, such that the error
originating from the top quark mass is not dominating the final error of the analysis.
Currently, this would call for a high precision on the value of the top mass, but not for a very high precision.
However, there are also reasons why a very high precision is desirable:
The top quark mass is close to the electro-weak symmetry breaking scale $v = 246 \; \mbox{GeV}$.
If there is new physics associated with electro-weak symmetry breaking, 
top quark physics is a place to look for.
New particles with masses above energies accessible with current collider experiments may nevertheless leave 
their traces in quantum corrections.
Therefore the combination of experimental precision measurements and theoretical precision calculations will be sensitive
to new physics at higher scales.
This is a very strong reason for a high precision determination of the top quark mass.
As a final reason let us also mention, that if we assume the Standard Model 
to be valid to very high scales (possibly as high as the Planck scale), the stability
of the electro-weak vacuum crucially depends on the precise numerical value of the top quark mass.

Let me also say from the very beginning that although I used the colloquial phrase ``the top quark mass'', 
there is nothing like ``the'' top quark mass.
Like any other parameter in the Lagrangian of the Standard Model, the top quark mass will be subject to renormalisation.
Like any other renormalised quantity, the renormalised top quark mass will depend on a chosen renormalisation scheme.
As there are several possible renormalisation schemes, there is more than one legitimate definition of a renormalised 
top quark mass.
In this talk I will discuss subtleties of some popular mass renormalisation schemes and the way they affect experimental measurements.

% -------------------------------------------------------------------------------------------
\section{Basic facts about the top quark}

The top quark is the heaviest elementary particle known up to today.
It has been discovered twenty years ago at the Tevatron \cite{Abe:1995hr,Abachi:1995iq}
and it is currently studied at the LHC.
The physics of the top quark is governed to a large extent by two essential numbers, the top quark mass
and the top quark width.
The current values of the top mass and the top width are \cite{Agashe:2014kda,ATLAS:2014wva}
\bq
\label{basic_numbers}
 m_t \;\; = \;\; 173.21 \pm 0.51 \pm 0.71 \; \mathrm{GeV},
 & &
 \Gamma_t \;\; = \;\; 2.1 \pm 0.5 \; \mathrm{GeV}.
\eq
With a mass of roughly $173 \; \mathrm{GeV}$ 
the top quark is heavier than all other known elementary particles.
This large mass sets also a hard scale.
From the top quark width one deduces immediately that the top quark lifetime ($\tau_t = \hbar / \Gamma_t$)
is shorter than the characteristic hadronisation time scale.
This implies that the top quark decays before it can form bound states.
Given the facts that the large top quark mass sets a hard scale and that top quarks do not hadronise it follows
that top quark physics is an ideal place for the application of perturbative QCD.

On the other hand it should not be forgotten that the top quark is like any other quark a colour-charged particle.
Furthermore, the top quark is like any other quark of the second or third generation an unstable particle.
These two facts imply that there is no asymptotic free top quark state in quantum field theory.
Although top quark physics is described mainly by perturbative QCD, 
one has to pay attention that non-perturbative effects 
-- originating from the fact that one deals with coloured and/or unstable particles --
do not enter from the back door.

A characteristic scale of non-perturbative effects is $\Lambda_{\mathrm{QCD}}$.
We can see from eq.~(\ref{basic_numbers}) that the error on the top quark mass is approaching 
${\mathcal O}(\Lambda_{\mathrm{QCD}})$.
This raises immediately the question if the top quark mass can be measured with a precision better than 
${\mathcal O}(\Lambda_{\mathrm{QCD}})$.
Of course, the top quark mass is determined from experimental measurements 
and it seems at first sight that reducing the error would just imply improving the experimental precision.
However, this is not the full story. 
Up to now there is no ``theory-free'' experimental determination of the top quark mass.
Experimental measurements of the top quark mass rely on theoretical input for example through the template method
or the matrix element method.
In this way theoretical uncertainties might enter the determination of the top quark mass.
There are now two possible scenarios, depending on the chosen mass definition.
In the first -- and not so favourable -- scenario the extraction of the top mass is limited 
by non-perturbative effects of order $\Lambda_{\mathrm{QCD}}$.
This means, that the precision on the top mass cannot be improved beyond ${\mathcal O}(\Lambda_{\mathrm{QCD}})$
by calculating perturbative higher-order corrections.
The pole mass definition is an example for this scenario.
In the second -- and more favourable -- scenario, one is not limited by non-perturbative effects and the precision
on the top mass can -- at least in principle -- be improved below ${\mathcal O}(\Lambda_{\mathrm{QCD}})$ by the inclusion
of perturbative higher-order corrections.
Short-distance mass definitions are examples of the second scenario.

% -------------------------------------------------------------------------------------------
\section{Basic facts about a fermion mass}

Let us now review the advantages and disadvantages of several mass definitions.
The starting point for a theoretical description is the Lagrange density, the relevant part reads
\bq
 {\cal L}_{\mathrm{fermion}} & = & \bar{\psi}_{\mathrm{bare}} \left( i D\!\!\!\!/ - m_{\mathrm{bare}} \right) \psi_{\mathrm{bare}}.
\eq
Beyond leading-order in perturbation theory loop diagrams have to be taken into account.
One of the simplest loop diagrams, which nevertheless allows us to discuss all relevant features, 
is the one-loop fermion self-energy:
\bq
\label{fermion_self_energy}
 - i \Sigma 
 \;\; = \;\;
\begin{minipage}{0.2\linewidth}
\includegraphics[scale=0.8]{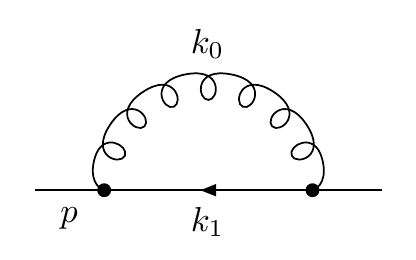}
\end{minipage}
 & = &
 \frac{g^2 C_F}{\mu^{D-4}} \int \frac{d^Dk}{(2\pi)^D} \; i \gamma_\rho \; \frac{i}{k\!\!\!/_1-m_{\mathrm{bare}}} \; i \gamma^\rho \; \frac{(-i)}{k_0^2}.
\eq
In four space-time dimensions the loop integral is divergent.
A convenient method of regularisation is the continuation of the number of space-time dimensions to $D=4-2\eps$,
the divergences will then show up as poles $1/\eps$.
Within dimensional regularisation one introduces in addition an arbitrary scale $\mu$ in order to
keep the mass dimension of the regulated expression to its four-dimensional value. 
The loop integral in eq.~(\ref{fermion_self_energy}) is easily computed and the result has with respect to the
spinor structure the form
\bq
 - i \Sigma 
 & = &
 -i \left( A p\!\!\!/ + B m_{\mathrm{bare}} \right).
\eq
Here, $A$ and $B$ are functions of $p^2$, $m^2_{\mathrm{bare}}$, $\mu^2$ and $\eps$.
As a function of $\eps$, the quantities $A$ and $B$ have a Laurent series expansion in $\eps$ starting with $\eps^{-1}$.
Iterations of self-energy insertions may be resummed similar to the resummation of a geometric series:
\bq
\label{resummed_propagator}
\begin{minipage}{0.1\linewidth}
\includegraphics[scale=0.8]{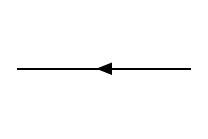}
\end{minipage}
+
\begin{minipage}{0.14\linewidth}
\includegraphics[scale=0.8]{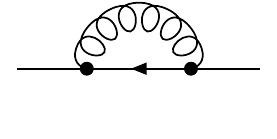}
\end{minipage}
+
\begin{minipage}{0.24\linewidth}
\includegraphics[scale=0.8]{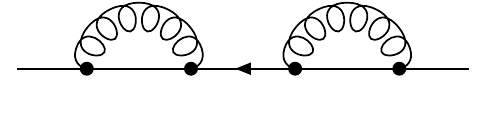}
\end{minipage}
 + 
 ...
 & = &
 \frac{i}{p\!\!\!/ -m_{\mathrm{bare}} - \Sigma}
 \\
 & = & 
 \frac{i (1+A)}{p\!\!\!/ - (1+A+B) m_{\mathrm{bare}} }
 + {\mathcal O}\left(\alpha_s^2\right).
 \nonumber
\eq
Renormalisation relates the bare quantities to the renormalised quantities.
We have to consider the quark field renormalisation and the mass renormalisation:
\bq
\label{renormalisation}
 \psi_{\mathrm{bare}} \;\; = \;\; \sqrt{Z_2} \; \psi_{\mathrm{renorm}},
 & &
 m_{\mathrm{bare}} \;\; = \;\; Z_m \; m_{\mathrm{renorm}}.
\eq
The quark field renormalisation allows us to absorb the divergences of the expression $(1+A)$ in the numerator of eq.~(\ref{resummed_propagator})
into $Z_2$,
the mass renormalisation allows us to absorb the divergences of $(1+A+B)$ in the denominator of eq.~(\ref{resummed_propagator}) into $Z_m$. 
It should be stressed that the renormalisation constants and hence the renormalised quantities 
depend on the renormalisation scheme.
In particular, the renormalised mass $m_{\mathrm{renorm}}$ depends on the renormalisation scheme.
All renormalisation schemes entail that they absorb the ultraviolet divergent terms.
Different renormalisation schemes differ in additional non-ultraviolet divergent terms.

% -------------------------------------------------------------------------------------------
\section{Implications on the precision for the top quark mass}

Let us now review different mass renormalisation schemes and its implications on the determination
of the renormalised top quark mass in a given scheme.

\subsection{The $\overline{\mathrm{MS}}$-scheme}

The $\overline{\mathrm{MS}}$-scheme absorbs by definition only the parts proportional to 
$\frac{1}{\eps} - \gamma_E + \ln(4\pi)$ and nothing else into the renormalisation constant $Z_m$.
The renormalised and resummed quark propagator of eq.~(\ref{resummed_propagator}) reads then
\bq
\label{MSbar_propagator}
 \frac{i}{ p\!\!\!/ - m_{\overline{\mathrm{MS}}} - (A+B)_{\mathrm{fin}} m_{\overline{\mathrm{MS}}} }
\eq
The essential properties of the $\overline{\mathrm{MS}}$-mass can already be deduced from eq.~(\ref{MSbar_propagator}):
Although not indicated explicitly, the $\overline{\mathrm{MS}}$-mass depends on the scale $\mu$, leading
to the concept of a running mass.
This follows from eq.~(\ref{renormalisation}):
The bare mass $m_{\mathrm{bare}}$ is of course scale-independent, while $Z_m$ and in consequence also $m_{\overline{\mathrm{MS}}}$ depend
on $\mu$.
Secondly, the presence of the finite terms $(A+B)_{\mathrm{fin}} m_{\overline{\mathrm{MS}}}$ in the denominator
shows, that the propagator does not have a pole at $p^2=m_{\overline{\mathrm{MS}}}^2$
and matrix elements do not factor at $p^2=m_{\overline{\mathrm{MS}}}^2$.
Thirdly, the extra terms $(A+B)_{\mathrm{fin}}$ in the denominator are not constant as a function of $p^2$, 
they vary with $p^2$. This implies that the propagator
in eq.~(\ref{MSbar_propagator}) will not yield a Breit-Wigner shape.
These properties should be kept in mind, when performing an analysis based on the $\overline{\mathrm{MS}}$-mass.

The $\overline{\mathrm{MS}}$-mass $m_{\overline{\mathrm{MS}}}$ is an example of a short-distance mass, meaning that the 
mass definition is not affected by long-distance non-perturbative effects.
The $\overline{\mathrm{MS}}$-mass can be extracted from 
an infrared-safe observable \cite{Langenfeld:2009wd,Dowling:2013baa}
for a process like $p p \rightarrow l \bar{\nu} j j b \bar{b}$ at high energies by comparing for example
$\sigma_{\mathrm{exp}}$ with $\sigma_{\mathrm{theo}}\left( m_{\overline{\mathrm{MS}}} \right)$.
It should be stressed that the error of such a measurement is not affected by an ${\mathcal O}(\Lambda_{\mathrm{QCD}})$-barrier.
This is related to the fact that $\overline{\mathrm{MS}}$-mass is a short-distance mass.
On the theory side, the uncertainty can systematically be improved by the inclusion of higher-order corrections.
The current state-of-the-art are NNLO calculations \cite{Czakon:2013goa}
with NNLL resummation \cite{Moch:2008qy,Czakon:2009zw,Cacciari:2011hy,Kidonakis:2010dk,Ahrens:2010zv,Beneke:2011mq}
for $p p \rightarrow t \bar{t}$ 
and NLO calculations for the process $p p \rightarrow b \bar{b} W^+ W^-$
including top decays and non-factorisable corrections \cite{Denner:2010jp,Bevilacqua:2010qb}.
At present, the dominant sources for the error on the determination of the  
$\overline{\mathrm{MS}}$-mass from cross section measurements originates 
from uncertainties on $\alpha_s$, the parton distribution functions and experimental uncertainties.
None of those are specific to the chosen mass definition.
A useful quantity in this context is the sensitivity ${\mathcal S}$ defined by
\bq
 \left| \frac{\delta \sigma}{\sigma} \right|
 & = &
 {\mathcal S}
 \;
 \left| \frac{\delta m_{\overline{\mathrm{MS}}}}{m_{\overline{\mathrm{MS}}}} \right|.
\eq
For the determination of the top mass from the total cross section for $t \bar{t}$-production 
the sensitivity is ${\mathcal S} \approx 5$.
The current error on the determination of the $\overline{\mathrm{MS}}$-mass $m_{\overline{\mathrm{MS}}}$ along these lines
is about $2 \; \mathrm{GeV}$.

As a significant fraction of $t \bar{t}$-events actually are accompanied by additional jets, also the process
$p p \rightarrow t \bar{t} + \mbox{jet}$ is of interest.
For this process NLO calculations are available \cite{Dittmaier:2007wz,Dittmaier:2008uj,Melnikov:2010iu,Melnikov:2011qx}.
Differential distributions for this process show a sensitivity in the range ${\mathcal S} \approx 10 ... 20$ and have
therefore the potential for a more precise extraction of the top mass \cite{Alioli:2013mxa}.

\subsection{The on-shell-scheme}

\begin{figure}
\begin{center}
\label{self_energy_insertions}
\includegraphics[scale=0.8]{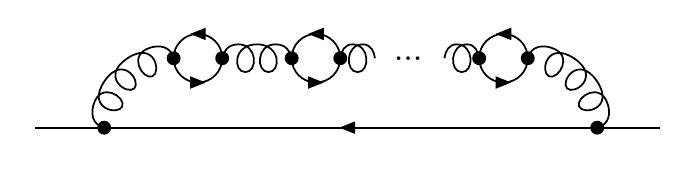}
\caption{Self-energy insertions on the gluon line leading to the renormalon ambiguity.}
\end{center}
\end{figure}

In the on-shell scheme the mass renormalisation constant $Z_m$ is defined in such a way that the
propagator has a pole at $m_{\mathrm{pole}}$,
(and $m_{\mathrm{pole}}$ is therefore called the pole mass).
The renormalised and resummed quark propagator is then by definition
\bq
\label{pole_mass_propagator}
 \frac{i}{ p\!\!\!/ - m_{\mathrm{pole}} }.
\eq
The pole mass $m_{\mathrm{pole}}$ includes the width and is therefore a complex quantity.
The pole mass has the advantage that matrix elements factor at $p^2 = m_{\mathrm{pole}}^2$ and that
the propagator of eq.~(\ref{pole_mass_propagator}) leads to a Breit-Wigner shape.
However, there is a major disadvantage: 
The pole mass is not a short-distance mass and sensitive to long-distance non-perturbative effects.
In the on-shell scheme, the renormalisation constant $Z_m$ contains contributions from all momentum scales, 
not just the ultraviolet region. It can be shown that in higher order in perturbation theory subsets of diagrams like the one shown in fig.~\ref{self_energy_insertions}
are dominated by the infrared region. 
The renormalised light fermion insertions are given by
\bq
 - \frac{2}{3} N_f \frac{\alpha_s}{4\pi} \left[ \ln\left(\frac{-k^2}{\mu^2}\right) - \frac{5}{3} \right],
\eq
with $k$ being the gluon momentum, which still needs to be integrated over. Due to the logarithm the ultraviolet and the infrared region are
enhanced.
A power series is Borel-summable if the Borel transform has no singularities on the real positive axis and does not increase too rapidly
at positive infinity.
With the replacement $\beta_{0,N_f} \rightarrow \beta_0$ one finds that the ultraviolet region leads to (non-critical) poles along
the negative real axis, while the infrared region leads to poles along the positive real axis.
Therefore this subset of diagrams is not Borel-summable and
the full perturbative series can only be summed up to an infrared renormalon ambiguity.
The renormalon ambiguity is of 
${\cal O}\left(\Lambda_{\mathrm{QCD}}\right)$ \cite{Bigi:1994em,Beneke:1994sw,Beneke:1994rs,Smith:1996xz}.
This ambiguity limits the precision by which the pole mass can be extracted from experiment.

In perturbation theory one can convert between different different renormalisation schemes.
We therefore have a relation between the $\overline{\mathrm{MS}}$-mass and the pole mass, 
which with the notation $\bar{m}=m_{\overline{\mathrm{MS}}}(\mu=m_{\overline{\mathrm{MS}}})$ reads
\bq
\label{conversion_formula}
 m_{\mathrm{pole}}
 & = &
 \bar{m}
 \times 
 \left[ 
         1 
         + c_1 \frac{\alpha_s(\bar{m})}{\pi}
         + c_2 \left( \frac{\alpha_s(\bar{m})}{\pi} \right)^2
         + c_3 \left( \frac{\alpha_s(\bar{m})}{\pi} \right)^3
         + c_4 \left( \frac{\alpha_s(\bar{m})}{\pi} \right)^4
         + ...
 \right].
\eq
The coefficients are known to four-loop order, the last coefficient $c_4$ was computed quite recently \cite{Chetyrkin:1999qi,Melnikov:2000qh,Marquard:2015qpa}.
Numerically, we have for the top quark:
\bq
 m_{\mathrm{pole}}
 & = &
 \bar{m}
 \times 
 \left[ 
         1 
         + 0.046
         + 0.010
         + 0.003
         + 0.001
         + ...
 \right].
\eq
The perturbative series appearing on the right-hand side of eq.~(\ref{conversion_formula})
is again only an asymptotic series and has an renormalon ambiguity as well.
This is clear from the fact that $\bar{m}$ is free of renormalon ambiguities, while $m_{\mathrm{pole}}$ on the left-hand side
suffers from a renormalon ambiguity.

Crude estimates of the renormalon ambiguity may be either obtained from renormalon-based calculations \cite{Beneke:1994sw}, yielding
\bq
 \delta m_{\mathrm{pole}} 
 & \approx &
 C_F \frac{2\pi}{\beta_0} e^{\frac{5}{6}} \Lambda_{\mathrm{QCD}} \left( \ln \frac{\bar{m}^2}{\Lambda_{\mathrm{QCD}}^2} \right)^{-\frac{\beta_1}{2\beta_0^2}}
 \;\; \approx \;\;
 {\mathcal O}\left( 300 \; \mathrm{MeV} \right),
\eq
with $\beta_0=11-2 N_f/3$ and $\beta_1=102-38 N_f/3$, or 
from the last known term in the conversion formula in eq.~(\ref{conversion_formula}).
The latter gives
\bq
 \delta m_{\mathrm{pole}} 
 & \approx &
 c_4 \; \bar{m} \left( \frac{\alpha_s(\bar{m})}{\pi} \right)^4
 \;\; \approx \;\;
 {\mathcal O}\left( 200 \; \mathrm{MeV} \right).
\eq
Let us stress that both numbers are just crude estimates.
Let us also note that the spread of two (or more) ad-hoc non-perturbative models might not reflect the true uncertainty
from non-perturbative effects.

Let us finally mention that the top width is not affected by a renormalon ambiguity, when expressed in terms of
a short-distance mass \cite{Bigi:1994em,Smith:1996xz}.

\subsection{The $\mathrm{MSR}$-scheme}

We have seen that the pole mass is ambiguous by an amount of order ${\cal O}\left(\Lambda_{\mathrm{QCD}}\right)$.
But on the other hand the measurement of the peak position of the decay products of the top quark
is an experimental observable.
This brings us to the question, if one can translate a measurement of the peak position into a
theoretical well defined short-distance mass.
As experimentalists can measure many things to high precision 
(like for example the average number of pions in $p p$ collisions),
the question is if and how a measured quantity can be related to a quantity depending only on short-distance physics
(the average number of pions is not a short-distance quantity).
Before answering this question, let us analyse the problem in more detail.
We should first find out, which scales are involved. In a second step we have to address the question
on how to define a short-distance mass at a given scale.
In the final step we then tackle the issue on how to translate the measurement into a short-distance mass.

Let us start with the involved scales.
Effective theories are the appropriate tool to describe the relevant degrees of freedom at a given scale $\mu$.
Evolution operators allow us to move from a scale $\mu_1$ to a scale $\mu_2$.
The evolution operators sum up large logarithms and avoid in this way large logarithms, which may otherwise spoil
a perturbative expansion.
Applied to the top mass, this has been analysed in detail for 
top pair production in electron-positron annihilation \cite{Fleming:2007qr,Fleming:2007xt} and similar results
are expected to hold for $p p$-collisions \cite{Hoang:2008xm}.
The relevant scales are the centre-of-mass energy $Q$, the top mass $m_t$, the top width $\Gamma_t$
and $\Lambda_{\mathrm{QCD}}$.
These scales are ordered as
\bq
 \Lambda_{\mathrm{QCD}} < \Gamma_t  < m_t < Q.
\eq
In the range $[m_t,Q]$ physics is described by QCD, while in the range $[\Gamma_t,m_t]$
the appropriate description is in terms of soft-collinear effective theory (SCET).
At even lower scales $[\Lambda_{\mathrm{QCD}},\Gamma_t]$ one uses a version of heavy quark effective theory adapted to
top quarks (top-HQET).
The relevant matrix elements for the various effective theories
are the hard function, the jet function and the soft function, respectively.
The impact on the invariant mass distribution from the various scales is as follows:
Scales in the range $[\Gamma_t,m_t]$ affect mainly the norm of the distribution.
The change in normalisation depends on $m_t$.
Scales from the range $[\Gamma_t,m_t]$ have an impact on the shape and the position of the peak and these effects
depend again on $m_t$.
For the low scales from the range $[\Lambda_{\mathrm{QCD}},\Gamma_t]$ one finds that these scales influence
as well the shape and the position of the peak. However it is important to note that those effects
are independent of $m_t$.
The situation is summarised in table~\ref{table_1}.
\begin{table}
\caption{Summary of the relevant scales, the appropriate effective theories together with the relevant matrix elements.
Also indicated is the impact on the peak distribution and the dependence on the top mass.}
\label{table_1}
\vspace{0.4cm}
\begin{center}
\begin{tabular}{|l|l|l|l|l|}
\hline
 Scale & Effective & Matrix & Impact on invariant & Top mass \\
 & theory & elements & mass distribution & dependence \\
 \hline
 $Q ... m_t$ & QCD & hard function & norm of the distribution & depends on $m_t$ \\
 $m_t ... \Gamma_t$ & SCET & jet function & shape and position & depends on $m_t$ \\
 $\Gamma_t ... \Lambda_{QCD}$ & top-HQET & soft function & shape and position & independent of $m_t$ \\
\hline
\end{tabular}
\end{center}
\end{table}
Since the effects from the scales $[\Lambda_{\mathrm{QCD}},\Gamma_t]$ are independent of $m_t$, it follows
that we need a short-distance mass definition for scales down to $\Gamma_t$.
The basic idea for the construction of an appropriate short-distance mass is to remove contributions which would
give rise to the renormalon ambiguity.
This approach is taken from experience with bottomium physics, where short-distance masses
like the potential subtracted mass $m_{\mathrm{PS}}$ \cite{Beneke:1998rk}
or the $1\mathrm{S}$-mass $m_{1\mathrm{S}}$ \cite{Hoang:1998hm,Hoang:1999zc} have been considered.
For the top quark this will involve apart from the UV-renormalisation scale $\mu$ a second scale $R$.
The $\overline{\mathrm{MS}}$-mass is an example of a short-distance mass and we have $R=\bar{m}$ in this case.
The $\mathrm{MSR}$-mass \cite{Hoang:2008yj} is a two-scale generalisation with a UV-scale $\mu$ and an IR-scale $R$, such that
\bq
 m_{\mathrm{MSR}}\left(R=0\right)
 =
 m_{\mathrm{pole}},
 & &
 m_{\mathrm{MSR}}\left(R=\bar{m}\right)
 =
 \bar{m}.
\eq
We may think of a short-distance mass definition in the same way as we think about an infrared-safe jet definition.
A jet is defined by the specification of a jet algorithm (SISCone, $k_t$-algorithm, anti-$k_t$-algorithm, etc.)
and by a set of parameters associated to this algorithm ($R$, $f$, $n_{\mathrm{pass}}$, $y_{\mathrm{cut}}$, etc.).
In the same way a short-distance mass is defined by the specification of
a short-distance renormalisation scheme ($\overline{\mathrm{MS}}$-scheme, MSR-scheme, etc.)
and by a set of parameters associated to this renormalisation scheme ($\mu$, $R$, etc.).

As the soft function is independent of the top mass (and information on the soft function may be obtained from
massless jet distributions), the peak position of the top invariant mass distribution can be related 
to a short-distance mass at a scale of $\Gamma_t$.

We now discuss how a measurement of the top invariant mass distribution can be translated into a short-distance mass.
In the actual extraction of the top mass from experimental measurements theory sneaks in through the use
of the template method or the matrix element method.
For example, within the template method one generates first from Monte Carlo events for various values of $m_{\mathrm{MC}}$
and then determines the best fit to the experimental data.
The Monte Carlo mass $m_{\mathrm{MC}}$ is only implicitly defined through the program code of the Monte Carlo.
However, the factorisation of the effective field theory approach (hard function/jet function/soft function)
has an analogy in typical event generators (hard matrix element/parton shower/hadronisation) and the shower cut-off scale
is typically of the order of $\Gamma_t$ (this is a numerical coincidence).
Because the shower cut-off provides a strict infrared cut-off for long-distance effects, it can be argued that the Monte Carlo mass
$m_{\mathrm{MC}}$ is something like a low-scale short-distance mass.
Therefore a measurement based on the top invariant mass distribution determines the Monte Carlo mass $m_{\mathrm{MC}}$,
which is a short-distance mass defined implicitly through the program code of a specific Monte Carlo.
In principle we could convert this mass to any other mass definition,
but in the case at hand we are hampered by the fact that a precise definition of the Monte Carlo mass is not accessible.
Parametrising the ignorance of a precise definition of the Monte Carlo mass, Hoang and Stewart \cite{Hoang:2008xm} made
in a contribution to the Top Quark Physics workshop in 2008 a first estimate for the translation to the $\mathrm{MSR}$-mass:
\bq
 m_{\mathrm{MC}}
 & = &
 m_{\mathrm{MSR}}\left(R = 1 ... 9 \; \mathrm{GeV}\right).
\eq
The uncertainty in the infrared scale $R$ introduces 
an uncertainty of the order of $1 \; \mathrm{GeV}$ on the translation from the Monte Carlo mass
to the $\mathrm{MSR}$ mass.
Let us summarise: The Monte Carlo mass is a (not so well specified) short-distance mass and the translation from the Monte Carlo
mass to a theoretically well defined short-distance mass at a low scale is currently estimated to be of the order
of $1 \; \mathrm{GeV}$ \cite{Hoang:2008xm,Moch:2014tta}.

\subsection{Work to do}

There are ample opportunities to improve the current state of the art. They can be grouped into three categories.

First of all, the details related to factorisation and the various effective theories have only been worked out
for $t \bar{t}$-production in electron-positron annihilation.
This remains to be done for $p p$-collisions.
Although it is believed that the general picture will hold in $p p$-collisions as well, there are some modifications
related to initial state partons and phase space cuts imposed by the jet algorithm.
These issues were absent in the $e^+ e^-$-analysis: There are no initial state partons and the analysis was based on
hemisphere masses.

Secondly, it is worth studying the translation from the Monte Carlo mass to a well defined short-distance mass in more detail.
In particular, one should firmly establish that the shower cut-off effectively implements some short-distance mass.
In addition, the translation and the uncertainty from the Monte Carlo mass to a well defined short-distance mass can be improved.

Thirdly, it is worth a thought to envisage a dedicated event generator, based on a well defined short-distance top mass.
This would eliminate the translation from a Monte Carlo mass to a well defined short-distance and the corresponding
uncertainties from this part.
This might not be impossible. In fact, there are proposals in the literature in a context not specific to top physics
to go from effective theories like SCET
to exclusive event generators \cite{Bauer:2006mk,Bauer:2008qh,Bauer:2008qj}.

% -------------------------------------------------------------------------------------------
\section{Conclusions}

The precise numerical value of the top mass is essential for many analyses in high-energy precision physics.
With the ongoing LHC experiments the error on the top mass is approaching ${\mathcal O}(\Lambda_{\mathrm{QCD}})$.
At this precision, an adequate short-distance mass definition is mandatory.
The pole mass is not a short-distance mass and ambiguous by an amount of ${\mathcal O}(\Lambda_{\mathrm{QCD}})$.
The $\overline{\mathrm{MS}}$-mass is a short-distance mass and can be used at high scales.
The $\mathrm{MSR}$-mass is a generalisation of the $\overline{\mathrm{MS}}$-mass and can also be used as a short-distance
mass at lower scales.

As an outlook towards the future it is expected that a threshold scan at an $e^+ e^-$-machine will be able
to determine the top quark mass with a precision below $100 \; \mathrm{MeV}$.
Again, the use of an adequate short-distance mass like the potential subtracted mass or the $1\mathrm{S}$-mass is
mandatory.

% -------------------------------------------------------------------------------------------
\section*{References}

\bibliography{/home/stefanw/notes/biblio}
\ifthenelse{\boolean{arxiveversion}}
{
\bibliographystyle{/home/stefanw/latex-style/h-physrev5}
}
{
\bibliographystyle{/home/stefanw/drafts/moriond15/h-elsevier2-mod}
}

\end{document}